# Effects of Asymmetric Cooling and Surface Wettability on the Orientation of the Freezing Tip

**Anton Starostin**
Institute of Technical Chemistry, 3 Academician Korolev St, Perm, 614013, Russia, 0000-0001-7484-8774
**Vladimir Strelnikov**
Institute of Technical Chemistry, 3 Academician Korolev St, Perm, 614013, Russia, 0000-0003-2538-535X
**Leonid A. Dombrovsky**
Chemical Engineering Department, Faculty of Engineering, Ariel University, P.O.B. 3, 407000, Ariel, Israel,
X-BIO Institute, University of Tyumen, 6 Volodarskogo St, Tyumen, 625003, Russia,
Heat Transfer Department, Joint Institute for High Temperatures, 17A Krasnokazarmennaya St, Moscow, 111116, Russia, 0000-0002-6290-019X
**Shraga Shoval**
Department of Industrial Engineering and Management, Faculty of Engineering, Ariel University, P.O.B. 3, 407000, Ariel, Israel, 0000-0002-0582-4821
**Oleg Gendelman**
Faculty of Mechanical Engineering, Technion−Israel Institute of Technology, Haifa 3200003, Israel, 0000-0002-4750-2708
**\*Edward Bormashenko**
Chemical Engineering Department, Faculty of Engineering, Ariel University, P.O.B. 3, 407000, Ariel, Israel, 0000-0003-1356-2486
edward@ariel.ac.il , +972 39066134

**Abstract**

Freezing of water droplets placed on the bare and superhydrophobic surfaces of polymer wedges are studied both experimentally and computationally. Two-dimensional numerical calculations of the transient temperature field in a chilled polymer wedge show that the direction of heat flux from the droplet through the thermal contact region with the wedge differs significantly from the normal to the wedge surface. This is the physical cause of the recently observed asymmetric cooling of the droplet. A novel approximate computational model is proposed that takes into account the variable area of the water freezing front in the droplet. This model gives a quantitative estimate of the faster freezing of the droplet on the bare surface. The obtained numerical results agree with the data of laboratory experiments. The velocity of the crystallization front and the droplet deformation including the so-called freezing tip formation are monitored in the experiment. The direction of the freezing cone axis appears to be noticeably different for the cases of bare and superhydrophobic wedge surfaces. This deviation is explained by the fact that the direction of the freezing cone axis is controlled by the local direction of the heat flux. For a hydrophobic wedge surface, the deviation of the freezing tip from the vertical is smaller, because the reduced thermal contact area reduces the influence of the heat flux direction at the wedge surface.

KEYWORDS: Contact angle, heat transfer, hydrophobic interface, phase transition; freezing.

**Introduction**

Icing is a widespread process, which is crucially important for multiple applications, including civil enginnering[1], aircraft[2,3] and energy production industries.[4] Insights into the physics of the icing are also important for understanding ice friction.[5,6] The icing at the solid-vapor interface is a complex process involving various thermo-physical and interfacial phenomena, with a decisive role in heat and mass transfer.[7-11] Development of the anti-icing surfaces remains an actual problem of applied physics.[12,13] The present study is focused on the final stage of droplet freezing when the so-called "freezing tip" (or freezing cone) is formed.[7,14-23] The formation of the sharp freezing cone is a fascinating counterintuitive phenomenon. Sharp, pointy singularities occur rarely due to their high energy concentration. The phenomenon has been reasonably related to the expansion of water upon freezing.[15-17] However, a quantitative description of the effect is incomplete at the moment. The freezing tip shape demonstrates surprising insensitivity to the freezing conditions, the volume of the droplet, the temperature of the substrate and the cooling rate, and even to the coating of the droplet with oils or powders.[24, 25] It might be mistakenly concluded that the vertical orientation of the geometrical axis of the freezing tip is dictated by gravity, which should impose the cylindrical symmetry of the frozen droplet. However, it was recently shown that the spatial orientation of the freezing tip depend mainly on the heat flux direction at the contact area of the droplet with the wedge surface.[26] Strictly speaking, the latter statement should be clarified, since the important thing is the heat flux field inside the freezing droplet, which also depends on the wettability of the wedge surface. As far as we know, the influence of the wettability of the wedge studied in this paper has not been considered before.

**Experimental**

Distilled water droplets were placed on the surface of the PMMA wedge as shown in **Figure 1**. Asymmetric cooling of the droplet is achieved by using a wedge made of material with relatively low thermal conductivity (**Figures 1-2**). The wedge was manufactured from polymethylmethacrylate (PMMA), CAS: 9011-14-7; supplied by Jumei Acrylic Manufacturing Co. Ltd. China; thermal diffusivity of PMMA $\kappa = 0.11 \times 10^{-6} \frac{m^2}{s}$; apparent contact angle $\theta = 70.0 \pm 5.0^0$.

Bi-distilled water with the specific resistivity $\rho = 17.8 \pm 0.5$ MΩ × cm established at $t = 25^0C$ was used in the experiments. Droplets with the volume of $V = 10 \pm 0.1$ µl were placed on the PMMA wedge with the precise micro-syringe. The dimensions of the PMMA wedge and its location on the cooled metallic plate are depicted in **Figure 1**.

The upper surface of the wedge was horizontal as shown in **Figure 2**. Visualization of the cooling process was carried out with the goniometer CRUSS DSA-100, GmbH, equipped with the software CRUSS Advance. The distance between the goniometer and PMMA wedge was 15 cm. The shape of the droplet was monitored with the CCD camera with a resolution of $1920 \times 1200$ pixels. Experiments were repeated 5 times; high repeatability of experiments was registered. Cooling of the polymer wedge was carried out with thermo-electric modulus of power 95 W enabling maximal temperature change $\Delta t = 84^0C$. The cooling rate of the surface of the thermo-electric modulus was $20 \pm 1$ K/min. The temperature of the surface of this modulus was controlled with the infrared sensor and in parallel with K-type thermocouple. The initial temperature and humidity is the experimental cell shown in **Figure 2** were established as $t = 25 \pm 0.5^0C$ and $RH = 40 \pm 1\%$ respectively. Control of the temperature within the experimental cell was performed with two thermocouples, the first of which measured the temperature of the upper surface of the thermoelectric modulus and the second one measured the temperature of the edge surface at the distance of 15 mm from its edge as shown in **Figure 1**. Polymer wedge was fixed on the surface of the thermo-electric modulus with the thermal paste with thermal conductivity $k = 14.2 \frac{W}{m\,K}$. Spatial orientation of the thermo-electric modulus ensured horizontality of upper surface of the edge, as shown in **Figure 2**. For this purpose, the thermo-electric modulus was inclined by angle $\alpha$ (**Figure 2**).

In two separate experiments, the water droplet with volume $V = 10$ µl was placed on the bare surface of the wedge and on the same surface coated with a superhydrophobic layer. The superhydrophobic layer has been created as follows. For the hydrophobization of the $SiO_2$ particles (Aerosil 380), at the first stage 0.4 g of perfluorodecyltrichlorosilane (FDTS) was introduced into 50 ml of Hexane. Afterwards, 1 g of $SiO_2$ was added to the solution, and the resulting mass concentration of the solution was 3%). The suspension was homogenized with the ultrasonic disperser Bandelin Sonopuls HD during 2 min. The suspension was dried with the two-stage process: i) drying during one hour under the temperature of 50°C; ii) drying during 30 min under $t = 150^0C$. Superhydrophobic layer was manufactured as follows: 0.5 g of the hydrophobized particles of Aerosil 380 were introduced into the solution of polymethylmethacrylate (PPMA) in dichloroethane. The mass concentration of the solution was 1%. The mass concentration of the hydrophobized particles of Aerosil 380 was 6%. Distribution of the particles of the solution was performed with the ultrasonic disperser Bandelin Sonopuls HD 3200 (modulus KE 76) during 30 s (total energy of homogenization was 0.8 kJ). The wedges were coated with the solution and dried during 30 s under ambient conditions. The coating was immersed in 0.1% Hexane solution (by mass) of FDTS and dried under the temperature of $t = 60^0 \text{C}$ during one hour. The wettability of the coating was studied with goniometer (Kruss DSA-100). Deionized water droplets were placed on the coating with the precise micro-syringe under ambient conditions. Measurements were averaged over ten experiments. Apparent contact angles were established as $\theta = 156 \pm 2^0$ (as depicted in **Figure 3**) contact angle hysteresis defined at the difference between advancing and receding contact angles $\Delta\theta = \theta_A - \theta_R = 2^0$. High apparent contact angle and low contact angle hysteresis evidence the Cassie-Baxter air trapping wetting regime, inherent for the superhydrophobic coating[27,28].

**Results and discussion**

Let's start with the experimental results and then move on to the computational part of the work. The freezing of the droplet was observed *in situ*. The shape of the droplet and the motion of the freezing front were visualized. The typical time history of the droplet shape under its freezing is presented in **Figure 4**. The initial shape of the droplet is close to spherical, due to the fact that the Bond number, defined as

$$\text{Bo} = \frac{\rho g R^2}{\gamma} \quad (1)$$

where $\rho = 1.0 \times 10^3 \frac{\text{kg}}{\text{m}^3}$ and $\gamma \cong 70 - 75 \frac{\text{mJ}}{\text{m}^2}$ are the density and surface tension of water in the temperature interval of $t = 5 - 20^0 \text{C}$. For the droplet with radius $R \cong 1\text{mm}$ Eq. (1) yields $\text{Bo} \cong 0.13 - 0.14$. Thus, the effect of gravity on the water droplet shape is negligible.

One can observe that the geometrical axis of the tip deviates from the vertical direction. This is true for both bare PMMA wedge and the wedge coated with the superhydrophobic layer, as shown in **Figure 4**. Moreover, this deviation is smaller for the case of superhydrophobic wedge surface. The monitoring of the process includes the visualization of the motion of the crystallization front within the droplet, illustrated in **Figure 5.** To rationalize these observations, one can recall that the orientation of the freezing cone is governed by the vector of the heat flux within the cooled droplet (orientation of the axis is illustrated with the yellow arrow in **Figure 4**).

A steadily moving crystallization front observed in the experiment is relatively simple, because the droplet is cooled from below. The point is that the density of water changes non-monotonically with temperature and is maximal not at the phase change temperature $T_0 = 0°\text{C}$, but at the temperature $T_* = 4°\text{C}$, when the water density is equal to $\rho_w \approx 1000 \text{ kg/m}^3$. This physical feature of water determines the temperature regime of lakes in the cold season, including the period when the lake is covered with ice. Related interesting geophysical and climatic phenomena are investigated in modern limnology[29,30]. For the convenience of quantitative estimates, the approximate formula for the temperature dependence of water density was suggested in early work[31]:

$$\rho_w = \rho_w^* \times \{1 - \varphi \times (T - T_*)^2\}, \qquad \varphi = 8 \times 10^{-6} \text{ K}^{-2} \quad (2)$$

In the temperature range from 0 to 8°C, this formula has an error of no more than 4%.

The density of water increases as the temperature changes from $T_0$ to $T_*$. Therefore, the temperature stratification of water in the droplet is unstable, which leads to natural convection: colder water rises and relatively warm water flows down in the direction of the crystallization front. This vortex convective motion observed in the experiment intensifies heat transfer in the water layer. The Grashof number which quantifies the ratio of the buoyancy force to the viscous force is directly proportional to $\Delta\rho \times H^3 \sim \varphi(\Delta T)^2 R^3$, where $H$ is the current thickness of the water layer above the ice-water interface, $\Delta T$ is the temperature difference in this layer, and $R$ is the droplet radius. Obviously, the small values of $\varphi$ and $R$ indicate that the average velocity of the vortex motion of water in the droplet is very small (this was also observed in the experiment). Most likely, the resulting contribution of the natural convection to the combined heat transfer in the water layer in the droplet is insignificant. This assumption has been already used in modeling of the liquid marble freezing[24]. A comparison of the calculated and measured freezing times for marbles of the same size as the water droplets in this study showed the validity of this assumption. At the same time, the natural convection of water in the droplet is important because it leads to a stable crystallization front separating the ice layer from the water above it.

According to paper[26], the temperature field in the polymer wedge under the droplet is not symmetric with respect to the normal to wedge surface passing through the center of the droplet. This is illustrated in **Figure 6**, where the calculations of paper[26] are completed by the field of the temperature gradient modulus.

As in **Figures 1** and **2**, the cooling and freezing water droplet in **Figure 6** is colored blue. Of course, this does not mean that the temperature is constant over the volume of the droplet and corresponds to the temperature scale given on the upper panel of this figure. An approximate calculation of the temperature field in the droplet is given below.

Obviously, the temperature field formed in the wedge, dictates the motion of the crystallization front and the formation of the freezing tip. At the same time, the temperature field in a small droplet has no effect on the temperature of the wedge. Therefore, the results of calculations for the water droplet using the computational model discussed below will not be needed to refine the temperature of the wedge under the freezing droplet.

In the suggested approximate model for the droplet cooling, we assume that the above discussed asymmetry does not strongly affect the crystallization front motion in the droplet and the axisymmetric model can be used to estimate the velocity of the crystallization front. In addition, we consider a novel modified 1D heat conduction problem instead of the complicated 2D heat conduction calculations[24]. The latter means that the assumption of the flat crystallization front is used and, in general, the temperature in each cross-section of the droplet, i.e. in the radial direction, is assumed to be constant.

The heat transfer from the ambient air to the droplet surface can be estimated using the minimum Nusselt number value $\mathrm{Nu} = 2$ for the stable temperature stratification of the air over the cold surface of the wedge. The latter allows us to use the formula $h = k_{\mathrm{air}}/R$ for the heat transfer coefficient, where $k_{\mathrm{air}}$ is the thermal conductivity of air. Calculations have shown that this effect is very small and can be neglected. On the contrary, it is necessary to take into account the large latent heat $L$ of phase transition.

The above discussed simplifications allow us to suggest the modified 1D model for the transient temperature profile $T(t, \bar{z})$:

$$\rho c R^2 \bar{S}(\bar{z}) \frac{\partial T}{\partial t} = \frac{\partial}{\partial \bar{z}} \left( k \bar{S}(\bar{z}) \frac{\partial T}{\partial \bar{z}} \right), \qquad 0 < \bar{z} < \bar{z}_{\mathrm{m}} = 1 + \sqrt{1 - \bar{r}^2} \qquad (2\mathrm{a})$$

$$T(0, \bar{z}) = T_0 + (T_1 - T_0)\, \bar{z}/\bar{z}_{\mathrm{m}}, \quad T(t, 0) = T_{\mathrm{s}}(t), \quad \left(\frac{\partial T}{\partial \bar{z}}\right)_{\bar{z}=\bar{z}_{\mathrm{m}}} = 0 \qquad (2\mathrm{b})$$

The geometric features of the problem are taken into account in Eq. (2a) with the use of the variable dimensionless cross-section area $\bar{S}$:

$$\bar{S}(\bar{z}) = \frac{S(t)}{S_0} = \frac{1-\left(\sqrt{1-\bar{r}^2}-\bar{z}\right)^2}{\bar{r}^2}, \quad S_0 = \pi r^2, \quad \bar{r} = \frac{r}{R}, \quad \bar{z} = \frac{z}{R} \qquad (3)$$

where $r$ is the radius of the thermal contact area of the droplet and the wedge and $z$ is the axial coordinate in the droplet measured from the contact surface. The word "thermal" is not occasional here because the thermal contact is greater than the mechanical contact is. The point is that the thermal resistance of a thin gas gap between the droplet and the substrate is very small and, as shown for the same working cell[24], does not prevent heat removal from the marble or water droplet. Of course, the change in values of $\rho c$ and $k$ at the moving ice-water interface is taken into account. According to ref. 24, the value of $T_1 = T_*$ was used in the initial condition. The first of the boundary conditions (2b) takes into account a decrease of the surface temperature of the wedge during the droplet crystallization[26]. The numerical solution to the transient conduction problem (2) was obtained using the traditional implicit finite-difference scheme[24,26].

There is no a special term in Eq. (3) corresponding to the heat source due to the water solidification. Instead, we use an equivalent additional heat capacity, $\Delta c$, in a narrow temperature range of $T_0 - \Delta T < T < T_0 + \Delta T$ for physically similar melting and solidification problems[32,33]:

$$\Delta c = \frac{L}{\Delta T}\left(1 - \frac{|T_0 - T|}{\Delta T}\right) \qquad (4)$$

A correct choice of $\Delta T$ value is determined by the interval of the computational grid and the time step of the numerical procedure. Some results of calculations are presented in **Figure 7**. The kink in the temperature profiles in **Figure 7a** corresponds to a strong change in thermal conductivity at the ice-water interface. As one might expect, the droplet cooling and freezing are much faster in the case of a good thermal contact between the droplet and the wedge surface. This agrees well with the laboratory observations. The very strong acceleration of the crystallization front at the final stage of the process is explained by a sharp decrease in the surface area of the crystallization front. At this stage, there is a qualitative change associated with the rapid release of the latent heat of the phase transition and the formation of a freezing tip.

The experimental findings also directly point on the effect of the superhydrophobicity. The freezing time at the bare surface was about $30 \pm 2$ s, for the superhydrophobic surface – $60 \pm 2$ s. This noticeable discrepancy points on a substantial difference in the thermal contact conditions. With a smaller contact area, the heat flux direction at the thermal contact area of the droplet and the wedge has smaller effect on the symmetry breaking in the freezing droplet; therefore, the observed deviation of the freezing tip axis from the vertical also decreases.

The relatively long freezing time in the model calculations (**Figure 7 b**) is mainly explained by the fact that it is difficult to correctly specify the real area of thermal contact between the droplet and the wedge surface, and this area is larger than the area of mechanical contact. At the same time, there is a qualitative agreement between the calculated and experimental data.

**Conclusion**

A recent study of the role of asymmetric cooling of sessile droplets on the orientation of the freezing tip was generalized to include the effect of hydrophobic surface under the droplet. New experimental data also showed that the total freezing time is significantly reduced for hydrophilic surfaces. This effect is accompanied by a marked change in the direction of the freezing cone axis. These effects have obtained a clear physical explanation. It was shown that the new approximate computational method proposed in the paper gives qualitatively good results for the propagation of the freezing front in the droplet.

**Acknowledgments**

The reported study was funded by the Russian Foundation for Basic Research, project number 19-29-13026/19. The authors are thankful to Dr. I. Legchenkova for her kind help in preparing this manuscript.

Final version is published in Surface Innovations, 2023; https://doi.org/10.1680/jsuin.22.01081

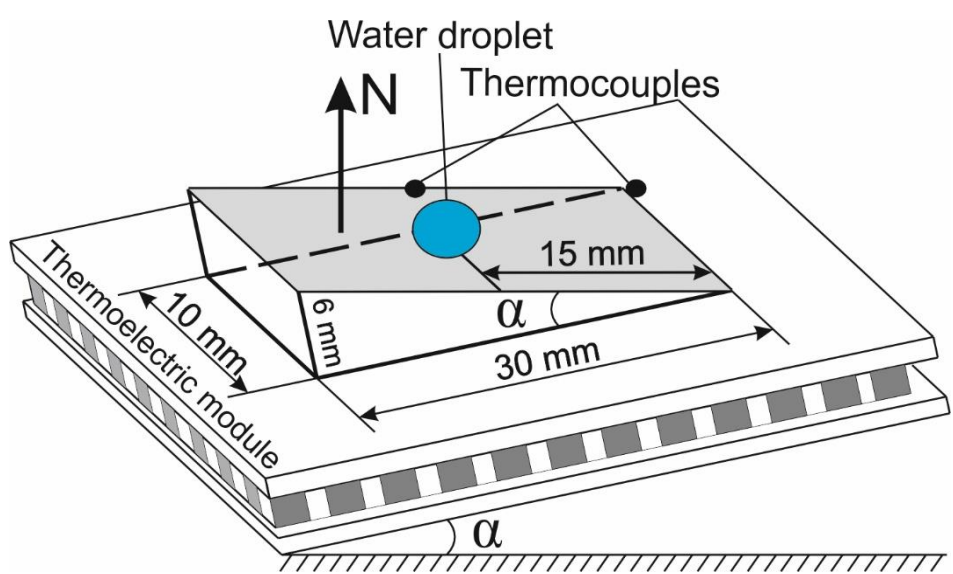
Figure 1. The scheme of the polymer wedge.

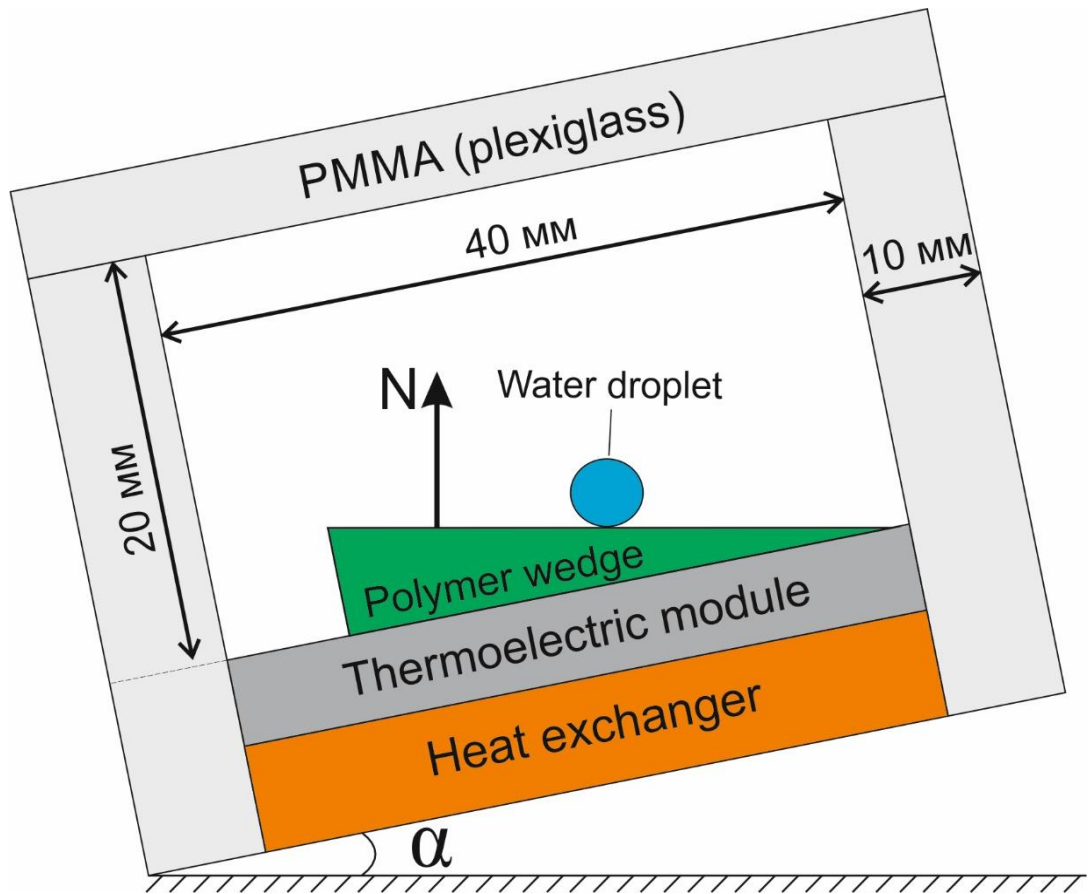

Figure 2. Scheme of the experimental edge used for the cooling of water droplets is depicted.

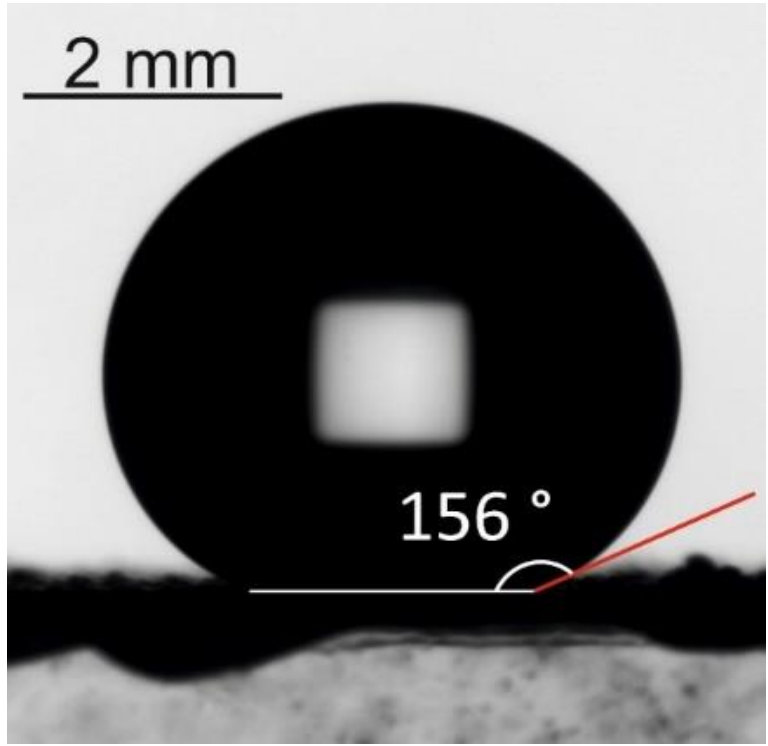

Figure 3. Apparent contact angle θ=156±2° on the superhydrophobic coating.

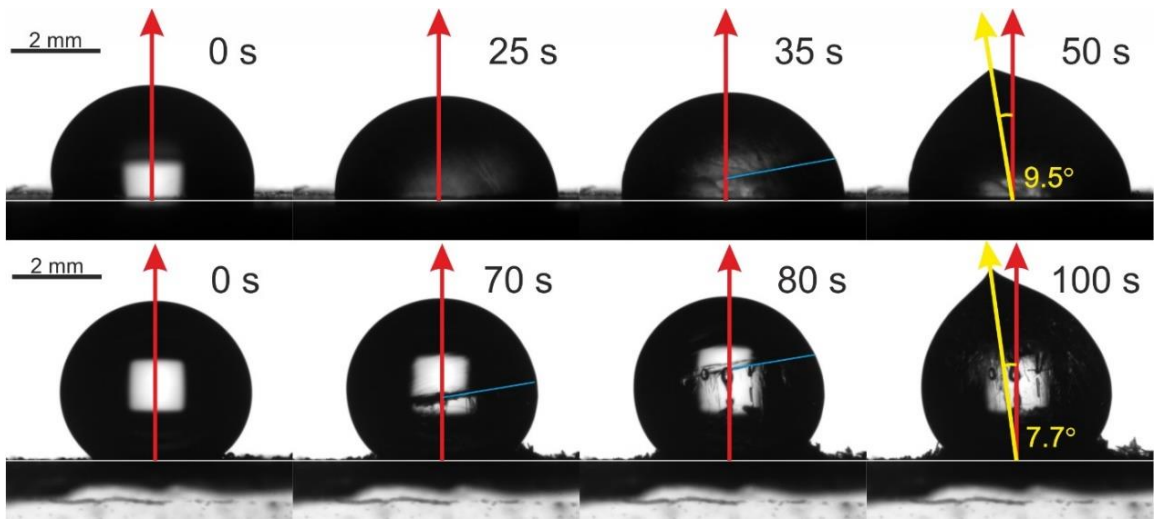

Figure 4. Freezing of the droplet placed on the bare PMMA wedge and the same wedge coated with the superhydrophobic layer. Red and yellow arrows depict the vertical direction and the axis of the freezing cone correspondingly.

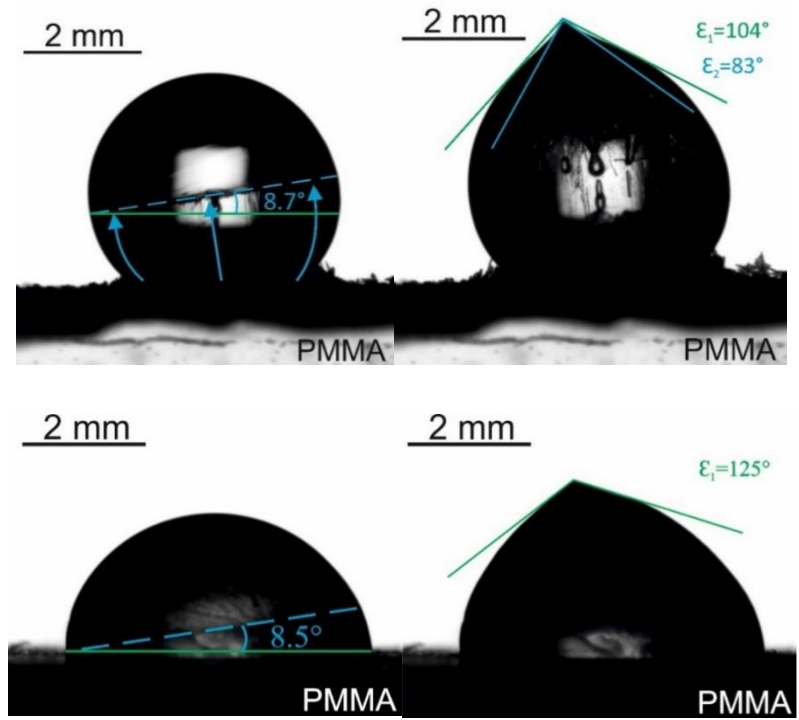

Figure 5. Movement of the crystallization front in the droplet placed on the surface of a polymer wedge with different hydrophobicity.

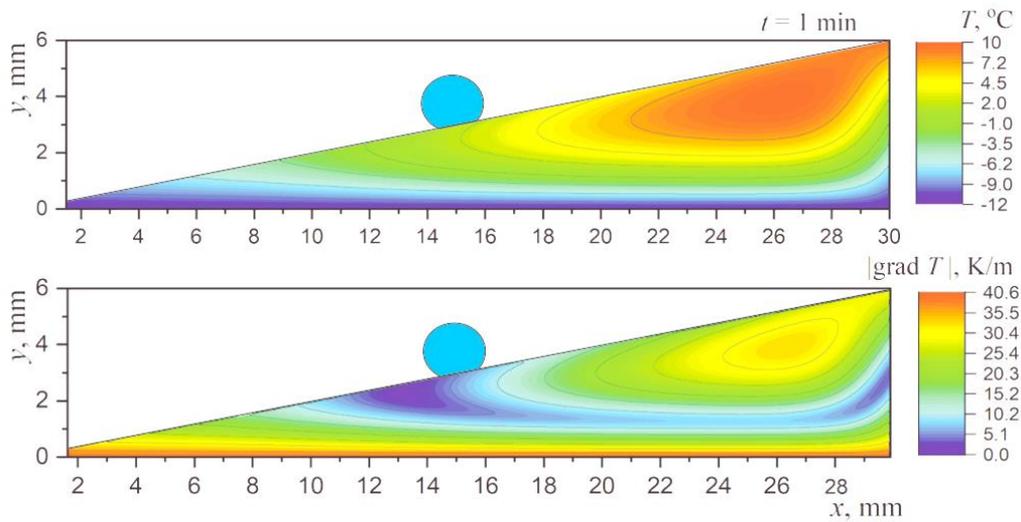

Figure 6. Temperature and temperature gradient modulus in the polymer wedge.

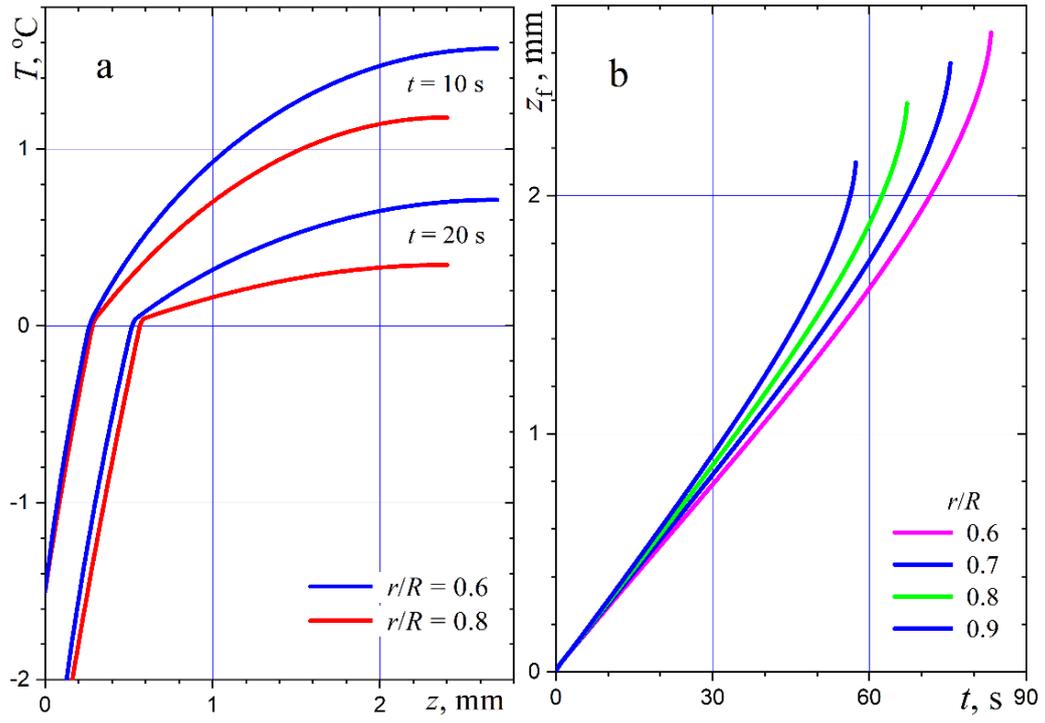

Figure 7. (a) Axial temperature profiles and (b) propagation of the crystallization front.